\documentclass[conference]{IEEEtran}
\usepackage[pdftex]{graphicx}
\usepackage{float}
\usepackage{latexsym}
\usepackage[cmex10]{amsmath}
\usepackage{amssymb}
\usepackage{amsmath}
\usepackage{paralist}
\usepackage{leftidx}
\usepackage{mathrsfs}
\usepackage{multirow}
\usepackage{slashbox}
\usepackage{amsthm}
\usepackage{background}
\usepackage{algorithm}
\usepackage{algorithmic}
\usepackage[norule]{footmisc}
\IEEEoverridecommandlockouts
\newcommand{\minitab}[2][l]{\begin{tabular}{#1}#2\end{tabular}}
\theoremstyle{definition}
\newtheorem{defn}{Definition}
\begin{document}
\SetBgContents{}
\title{QoS-Aware Joint Mode Selection and Channel Assignment for D2D Communications}
\author{\IEEEauthorblockN{Rui Wang$^*$, Jun Zhang$^*$, S.H. Song$^*$ and K. B. Letaief$^*$$^\dag$, \emph{Fellow, IEEE} }
\IEEEauthorblockA{$^*$Dept. of ECE, The Hong Kong University of Science and Technology, $^\dag$Hamad Bin Khalifa University, Doha, Qatar\\
Email: $^*$\{rwangae, eejzhang, eeshsong, eekhaled\}@ust.hk, $^\dag$kletaief@hbku.edu.qa}
\thanks{This work was supported by the Hong Kong Research Grant Council under Grant No. 610113.}
}

\maketitle
\begin{abstract}
  Underlaying device-to-device (D2D) communications to a cellular network is considered as a key technique to improve spectral efficiency in 5G networks. For such D2D systems, mode selection and resource allocation have been widely utilized for managing interference. However, previous works allowed at most one D2D link to access the same channel, while mode selection and resource allocation are typically separately designed. In this paper, we jointly optimize the mode selection and channel assignment in a cellular network with underlaying D2D communications, where multiple D2D links may share the same channel. Meanwhile, the QoS requirements for both cellular and D2D links are guaranteed, in terms of Signal-to-Interference-plus-Noise Ratio (SINR). We first propose an optimal dynamic programming (DP) algorithm, which provides a much lower computation complexity compared to exhaustive search and serves as the performance bench mark. A bipartite graph based greedy algorithm is then proposed to achieve a polynomial time complexity. Simulation results will demonstrate the advantage of allowing each channel to be accessed by multiple D2D links in dense D2D networks, as well as, the effectiveness of the proposed algorithms.
\end{abstract}
\IEEEpeerreviewmaketitle

\section{Introduction}

As a promising approach to improve network throughput, and reduce power consumption and delays, device-to-device (D2D) communications are proposed for next-generation cellular networks \cite{mag}. In cellular networks with underlaying D2D communications, proximity mobile users are allowed to communicate directly with each other under the control of base stations (BSs), rather than being forced to communicate via the BS \cite{design}. D2D communications bring new communication freedom, since, with D2D communications, mobile users may work in two modes, i.e., either the cellular mode or the D2D mode. Specifically, in the cellular mode, mobile users communicate via the BS, while in the D2D mode, mobile users communicate with each other directly. Thus, efficient mode selection schemes can help to improve the network throughput. However, without effcient interference management, severe interference will be generated for both cellular networks and D2D communications. Resource allocation is considered as an efficient approach to manage the interference and improve the network throughput \cite{survey}.

There have been lots of recent efforts on mode selection and resource allocation for D2D communications. Zhang \emph{et al.} \cite{nphard} proposed an interference-aware graph-based suboptimal algorithm to solve the channel assignment problem. Considering QoS requirements for both the cellular and D2D links, Feng \emph{et al.} \cite{one2one} proposed an algorithm based on maximum weighted bipartite matching which can get the optimal resource allocation scheme. In \cite{doppler2010mode}, the mode selection for a D2D link was investigated in a single cell system with one D2D link and one cellular user.

Recent works started to consider joint optimization of the mode selection and resource allocation for D2D communications. The authors in \cite{2link} obtained the optimal resource sharing and mode selection strategy with one cellular link and one D2D link. The case where multiple D2D links and cellular users are active in a single cell was considered in \cite{modepower}, where a distributed algorithm was proposed to manage the interference. In \cite{modepower}, each channel can be accessed by only one link. Considering the case where each cellular link can share the channel with at most one D2D link, authors in \cite{one2onemode} proposed three algorithms that can obtain the near optimal performance in different scenarios. However, in these works, it is assumed that at most one D2D link can access one channel, which fails to utilize the spectrum efficiently in a dense D2D network.

In this paper, we jointly optimize the mode selection and channel assignment in a cellular network with underlaying D2D communications in order to maximize the weighted sum-rate. Multiple D2D links are allowed to access the same channel, and both the cellular and D2D links are guaranteed to meet their QoS requirements. We first propose a dynamic programming (DP) algorithm to achieve the optimal joint mode selection and channel assignment, which can serve as the performance benchmark. Even though the DP algorithm has an exponentially increasing complexity, it is much more efficient than exhaustive search. To further reduce the complexity to polynomial time, we also propose a suboptimal algorithm, which firstly gets the channel assignment for cellular users based on the associated bipartite graph, and then, uses a greedy approach to obtain the joint mode selection and channel assignment for D2D links. Simulation results will validate the effectiveness of our proposed bipartite graph based greedy algorithm, and show that, the weighted sum-rate is significantly improved by jointly optimizing the mode selection and resource allocation. Moreover, we also illustrated the advantage of allowing multiple D2D links to share the same channel.

\section{System Model and Problem Formulation}
In this section, the system model will be firstly introduced, and then the joint channel assignment and mode selection problem will be formulated.

\subsection{System Model}
\begin{figure}[!t]
  \centering
  \includegraphics[width=2.5in]{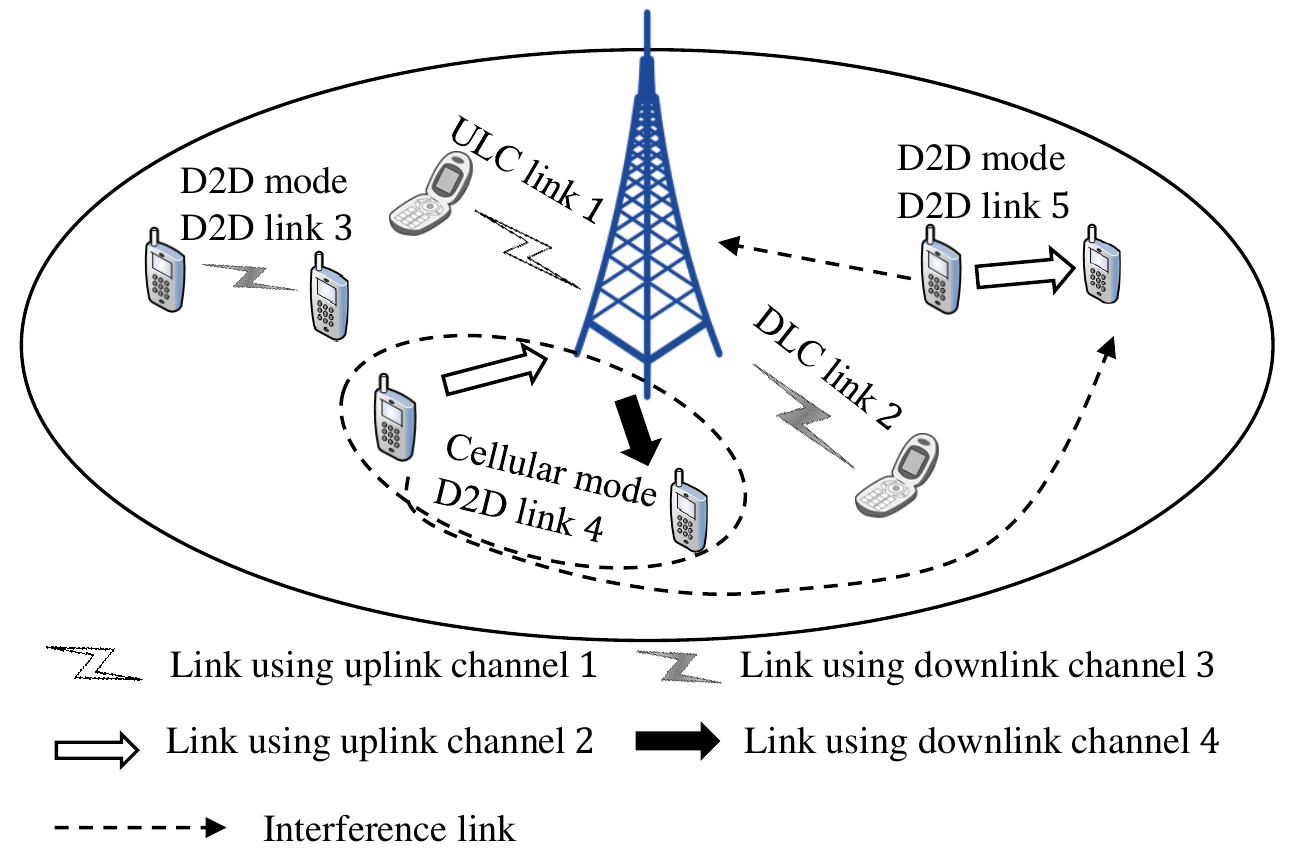}
  \caption{A sample network with two cellular links and three D2D links.}
  \label{model}
\end{figure}
A cellular network including $N_{uc}$ uplink cellular (ULC) users, $N_{dc}$ downlink cellular (DLC) users and $N_d$ D2D links is considered. The index sets of the ULC links, the DLC links and the D2D links are denoted as $\mathcal{C}_u = \{1,2,...,N_{uc}\}$, $\mathcal{C}_d = \{N_{uc}+1,N_{uc}+2,...,N_{uc}+N_{dc}\}$ and $\mathcal{D} = \{N_{uc}+N_{dc}+1,N_{uc}+N_{dc}+2,...,N_{uc}+N_{dc}+N_d\}$, respectively. Correspondingly, the index sets of the $N_c \triangleq N_{uc}+N_{dc}$ cellular links and $N \triangleq N_{c}+N_{d}$ communication links are denoted as $\mathcal{C} = \mathcal{C}_u \cup \mathcal{C}_d  = \{1,2,...,N_{c}\}$ and $\mathcal{S} = \mathcal{C} \cup \mathcal{D}  = \{1,2,...,N\}$, respectively. As in an LTE network \cite{LTE}, OFDMA is applied and there are $M_u$ uplink channels and $M_d$ downlink channels, whose index sets are denoted as $\mathcal{CH}_u = \{1,2,...,M_u\}$ and $\mathcal{CH}_d = \{M_u+1,M_u+2,...,M_u+M_d\}$, respectively. Correspondingly, the index set of the $M \triangleq M_{u}+M_{d}$ channels is denoted as $\mathcal{CH} = \mathcal{CH}_u \cup \mathcal{CH}_d  = \{1,2,...,M\}$. The numbers of uplink and downlink channels are assumed to be no less than the numbers of uplink and downlink cellular links, respectively. A sample network is shown in Fig. \ref{model}.

It is assumed that each cellular link can access at most one channel. Each D2D link may work in the following two modes:
\begin{enumerate}
    \item{The D2D mode:} The D2D transmitter directly communicates with its corresponding receiver. One uplink or downlink channel will be allocated to the D2D link.
    \item{The cellular mode:} The D2D pair communicates via the BS. One uplink and one downlink channels will be allocated to the D2D link.
\end{enumerate}
Considering the priority of cellular links, we assume that each cellular link must access one channel, while some D2D links may not be admitted to use any channel. Correspondingly, D2D links are called as active D2D links while accessing some channel, otherwise they are called inactive D2D links. Assuming that link $j \in \mathcal{S}$, $z \in \mathcal{S}$ access channel $i \in \mathcal{CH}$, and D2D links are operating in the D2D mode, we shall use $g^D_{i,j}$ and $h^D_{i,z,j}$ to denote the channel gains of link $j$ and the interference link from link $z$ to link $j$, respectively. Assuming that link $j \in \mathcal{S}$, $z \in \mathcal{S}$ access channel $i \in \mathcal{CH}$, and D2D links are operating in the cellular mode, we shall use $g^C_{i,j}$ and $h^C_{i,z,j}$ to denote the channel gains of link $j$ and the interference link from link $z$ to link $j$, respectively. The BS is assumed to be able to acquire the full channel state information (CSI). Cellular link $j \in \mathcal{C}$ is assumed to transmit with power $p^C_j$. D2D link $j \in \mathcal{C}$ is assumed to transmit with power $p^C_j$ when operating in the cellular mode, and is assumed to transmit with power $p^D_j$ when operating in the D2D mode. We assume the Additive white Gaussian noise with zero mean and variance $\sigma^2$ at each receiver. All cellular and active D2D links are assumed to have their minimum SINR requirements.

\subsection{Problem Formulation}

In this paper, we investigate the optimal joint channel assignment and mode selection to maximize the weighted sum-rate of the system while guaranteeing the QoS requirements for both cellular and D2D links. The channel assignment is denoted as $\rho_{i,j}$, with $i \in \mathcal{CH}$ and $j \in \mathcal{S}$, which is defined as

\begin{equation} \label{rho}
\rho_{i,j}=\left\{
                \begin{array}{ll}
                1 & \text{if channel } i \text{ is assigned to link } j,\\
                0& \text{otherwise}.
                \end{array}
                \right.
\end{equation}
Let $x_j$, with $j \in \mathcal{D}$, denote the mode selection for D2D links, and $x_j$, with $j \in \mathcal{C}$, always equals $1$. Then, $x_j$, with $j \in \mathcal{S}$, can be specified as
\begin{equation} \label{x}
x_j=
\begin{cases}
1 & j \in \mathcal{C} \text{, or } j \in \mathcal{D} \text{ and D2D link $j$ uses cellular mode,}\\
0 & j \in \mathcal{D} \text{ and D2D link $j$ uses D2D mode.}\\
\end{cases}
\end{equation}
The set $\mathcal{L}^C_i$ includes all the cellular links and cellular mode D2D links accessing channel $i$, given by
\begin{equation} \label{li_c}
\mathcal{L}^C_i=\left\{j|j \in \mathcal{S} \text{, } \rho_{i,j}=1 \text{ and } x_j=1\right\}.
\end{equation}
The set $\mathcal{L}^D_i$ includes all the D2D mode D2D links accessing channel $i$, given by
\begin{equation} \label{li_d}
\mathcal{L}^D_i=\left\{j|j \in \mathcal{S} \text{, } \rho_{i,j}=1 \text{ and } x_j=0\right\}.
\end{equation}
Then, the received SINR of link $j$ using channel $i$ is
\begin{align}
\xi_{i,j} &\left( \mathcal{L}^C_i, \mathcal{L}^D_i, x_j \right) = \notag \\
&{\frac{x_j p^C_j h^C_{i,j}+(1-x_j) p^D_j h^D_{i,j}}
{{\sigma^2  + \sum\limits_{
z \in \mathcal{L}^C_i, z \neq j} {p^C_z h^C_{i,z,j} }+\sum\limits_{z \in \mathcal{L}^D_i, z \neq j} {p^D_z h^D_{i,z,j} } }}}.\label{SINR}
\end{align}
We consider the weighted sum-rate as the performance metric, where the priorities and fairness of different users can be adjusted by the weights, e.g., D2D users may have a lower priority than cellular users. The weighted sum-rate maximization problem can then be formulated as
\begin{align}
\mathop {\max }\limits_{\rho_{i,j}, x_j } &\sum\limits_{j \in \mathcal{S}} {w_j \log \left( {1 + \text{SINR}_j } \right) } ,\label{problem}\\
\text{s.t. } &\text{SINR}_j= \sum \limits_{i \in \mathcal{CH}} \rho_{i,j} \xi_{i,j}\left( \mathcal{L}^C_i,\mathcal{L}^D_i,x_j \right)  \geqslant \xi^{\text{min}}_{j} , \forall j \in \mathcal{C} \tag{\ref{problem}a}\\
&\text{SINR}_j= (1-x_j)\left(\sum \limits_{i \in \mathcal{CH}} \rho_{i,j} \xi_{i,j}\left( \mathcal{L}^C_i,\mathcal{L}^D_i,x_j \right) \right) \notag \\
&+ x_j \min \bigg( \sum \limits_{i \in \mathcal{CH}_u} \rho_{i,j} \xi_{i,j}\left( \mathcal{L}^C_i,\mathcal{L}^D_i,x_j \right),\notag \\
&\sum \limits_{i \in \mathcal{CH}_d} \rho_{i,j} \xi_{i,j}\left( \mathcal{L}^C_i,\mathcal{L}^D_i,x_j \right) \bigg)
\geqslant \xi^{\text{min}}_{j} , \notag \\
&\forall j \in \mathcal{D} \text{ and } \sum\limits_{i \in \mathcal{CH}} {\rho _{i,j} } \geqslant 1, \tag{\ref{problem}b} \\
&\sum\limits_{j \in \mathcal{S}} x_j {\rho _{i,j} } \leqslant 1, \forall i \in \mathcal{CH}, \tag{\ref{problem}c} 
\end{align}
\begin{align}
&\sum\limits_{i \in \mathcal{CH}} {\rho_{i,j} } = 1, \forall j \in \mathcal{C}, \tag{\ref{problem}d}\\
&(1-x_j)\left(\sum\limits_{i \in \mathcal{CH}} {\rho_{i,j} } \right) \leqslant 1, \forall j \in \mathcal{D}, \tag{\ref{problem}e}\\
&x_j \left(\sum\limits_{i \in \mathcal{CH}_e} {\rho_{i,j} } \right) \leqslant 1, \forall j \in \mathcal{D}, e \in \{u,d\}, \tag{\ref{problem}f}\\
&\sum\limits_{i \in \mathcal{CH}_{e_1}} {\rho_{i,j} } = 0, \forall j \in \mathcal{C}_{e_2}, (e_1,e_2) \in \{(u,d),(d,u)\}, \tag{\ref{problem}g}
\end{align}
where $\xi^{\text{min}}_{j}$ represents the minimum SINR requirement of link $j$, and $w_j$ denotes the weight of link $j$. The QoS requirements for both cellular and active D2D links are guaranteed with constraints (\ref{problem}a) and (\ref{problem}b). Constraint (\ref{problem}c) ensures different cellular links and cellular mode D2D links cannot access the same channel. Constraints (\ref{problem}d) and (\ref{problem}e) imply that each cellular link can access one channel and each D2D mode D2D link cannot access more than one channel, respectively. Constraint (\ref{problem}f) ensures that each cellular mode D2D link can access at most one uplink channel and one downlink channel. Constraint (\ref{problem}g) ensures that uplink/downlink cellular links can only access uplink/downlink channels.

\section{Optimal joint mode selection and channel assignment}
The problem formulated in (\ref{problem}) is a mixed integer programming (MIP) problem and is NP-hard. In this section, a DP algorithm, with much lower complexity than exhaustive search, is proposed to find the optimal joint mode selection and channel assignment.

\subsection{Optimal DP Algorithm}
As an efficient approach to handle non-continuous solution spaces, the DP algorithm divides the original problem into multiple stages and associates each stage with multiple states \cite{dpint}. By achieving the optimal solution at the beginning stage and finding a recursive relationship of the optimal solutions at one stage and previous stages, the optimal solution of the original problem can be constructed. In the following, the stages, the states and the recursive relationship in our algorithm will be identified.

In our proposed DP algorithm, the stage represents the number of arranged uplink channels, and at each stage, a particular uplink channel is assigned. Meanwhile, the states include different subsets of $\mathcal{S}$ and different subsets of $\mathcal{CH}_d$, representing a particular group of links and a particular group of downlink channels, respectively. Specifically, at the $k$-th stage, $0 \leqslant k \leqslant M_u$, associated with state $\left\{\mathcal{J} \subseteq \mathcal{S}, \mathcal{Z} \subseteq \mathcal{CH}_d \right\}$, we need to maximize the weighted sum-rate of links in $\mathcal{J}$ sharing the uplink channels in $\mathcal{CH}^{k}  \triangleq \{i|1 \leqslant i \leqslant k \}$ and downlink channels in $\mathcal{Z}$. Correspondingly, the maximum weighted sum-rate is denoted as $OPT_{k,\mathcal{J},\mathcal{Z}}$, the optimal channel assignment of link $j \in \mathcal{S}$ is denoted as $\left(\rho_{k,\mathcal{J},\mathcal{Z}} \right)_{i,j}$, which is defined similarly as (\ref{rho}), and the optimal mode selection of $j \in \mathcal{S}$ is denoted as $\left(x_{k,\mathcal{J},\mathcal{Z}} \right)_{j}$, which is defined similarly as (\ref{x}). 

At the $0$-th stage, since no uplink channel is assigned, no D2D link can work in the cellular mode. Then, the problem becomes an optimal channel assignment problem, which can be solved by the DP algorithm proposed in \cite{conf}. Then, we can get the optimum value $OPT_{0,\mathcal{J},\mathcal{Z}}$, where $\mathcal{J} \subseteq \mathcal{C}_d \cup \mathcal{D}$ and $\mathcal{Z} \subseteq \mathcal{CH}_u$, and the corresponding channel assignment $\left(\rho_{0,\mathcal{J},\mathcal{Z}} \right)_{i,j}$. The mode selection at the $0$-th stage, i.e., $\left(x_{0,\mathcal{J},\mathcal{Z}} \right)_{j}$, equals $1$ for cellular links and $0$ for D2D links.

Then, to find the recursive relationship, we will show that the problem of finding the optimum value $OPT_{k,\mathcal{J},\mathcal{Z}}$ can be transferred to the problem of finding the optimum values of previous stages. In order to transform the problem, the following decisions should be made in sequence:
\begin{enumerate}
  \item selecting the set of cellular links and cellular mode D2D links sharing uplink channel $k$, denoted as $\mathcal{X}^{CS} \subseteq \mathcal{J}$,
  \item selecting the set of D2D mode D2D links sharing uplink channel $k$, denoted as $\mathcal{X}^{DSu}\subseteq \mathcal{J}$,
  \item selecting a downlink channel $d$, which is chosen in $\mathcal{Z}$ if $\mathcal{X}^{CS} \cap \mathcal{D} =1$, and equals $0$ otherwise,
  \item selecting the set of D2D mode D2D links sharing downlink channel $d$, denoted as $\mathcal{X}^{DSd}\subseteq \mathcal{J}$, which is a null set if $d=0$,
\end{enumerate}
and for simplification, we denote $\mathcal{X}^{All}=\mathcal{X}^{CS} \cup \mathcal{X}^{DSu}\cup \mathcal{X}^{DSd}$ as all the links using channel $k$ and $d$, and $\mathcal{X}=\{\mathcal{X}^{CS},\mathcal{X}^{DS}_u,d, \mathcal{X}^{DSd}\}$ as an element including all the decisions. Consequently, the uplink channels in $\mathcal{CH}^{k-1}$ and the downlink channels in $\mathcal{Z}-\{d\}$ can only be assigned to the links in $\left( \mathcal{J}-\mathcal{X}^{All} \right)$. Then, the problem can be transferred to finding the optimum value $OPT_{ k-1 ,  \mathcal{J}- \mathcal{X}^{All}, \mathcal{Z}-\{d\}}$. After searching for all possible selections of $\mathcal{X}$, the optimum value $OPT_{k,\mathcal{J},\mathcal{Z}}$, $1 \leqslant k \leqslant M$, can be achieved according to the following recursive relationship
\begin{align}
OP&T_{k,\mathcal{J},\mathcal{Z}} =  \notag \\
\mathop {\max } \limits_{ \mathcal{X} } & \big[ U \left( k,\mathcal{X} \right)
+ OPT_{k-1, \mathcal{J} - \mathcal{X}^{All}, \mathcal{Z}-\{d\}} \big] ,\label{rec} \\
\text{s.t. } & \mathcal{X}^{CS},\mathcal{X}^{DSu},\mathcal{X}^{DSd} \subset \mathcal{J}, \tag{\ref{rec}a} \\
& \mathcal{X}^{CS} \cap \mathcal{X}^{DSu}=\mathcal{X}^{CS} \cap \mathcal{X}^{DSd}=\mathcal{X}^{DSu} \cap \mathcal{X}^{DSd}=\varnothing \tag{\ref{rec}b} \\
&\xi_{k,j} \left( \mathcal{X^{CH}},\mathcal{X}^{DSu},1 \right) \geqslant \xi^{\text{min}}_{j}, \forall j \in \mathcal{X}^{CS} \tag{\ref{rec}c} \\
&\xi_{k,j} \left( \mathcal{X^{CH}},\mathcal{X}^{DSu},0 \right) \geqslant \xi^{\text{min}}_{j}, \forall j \in \mathcal{X}^{DSu} \tag{\ref{rec}d} \\
&\xi_{d,j} \left( \mathcal{X^{CH}},\mathcal{X}^{DSd},1 \right) \geqslant \xi^{\text{min}}_{j}, \forall j \in \mathcal{X}^{CS} \cap \mathcal{D} \tag{\ref{rec}e} \\
&\xi_{d,j} \left( \mathcal{X^{CH}},\mathcal{X}^{DSd},0 \right) \geqslant \xi^{\text{min}}_{j}, \forall j \in \mathcal{X}^{DSd} \tag{\ref{rec}f} \\
&\left|\mathcal{X}^{CS} \right| \leqslant 1 \tag{\ref{rec}g} \\
&\left| \mathcal{C} \cap \mathcal{X}^{CS} \right| = 1, \text{ if } \left| \mathcal{CH}_u \cap \mathcal{CH}^k \right| \leqslant \left| \mathcal{C}_u \cap \mathcal{J} \right|  \tag{\ref{rec}h} \\
& |\mathcal{D} \cap \mathcal{X}^{CU}|=0, \text{ if } |\mathcal{C}_d \cap \mathcal{J}| =|\mathcal{Z}| \tag{\ref{rec}i} \\
& \mathcal{X}^{CU} \cap \mathcal{C}_d = \varnothing \tag{\ref{rec}j}
\end{align}
where $|\cdot|$ denotes the cardinality and $U \left( k,\mathcal{X} \right)$ is the weighted sum-rate of links in $\mathcal{X}^{All}$, which is given by
\begin{align}
U \left( k,\mathcal{X} \right) &= \sum\limits_{j \in \mathcal{X}^{CS} \cap \mathcal{C}} {w_j \log \left( {1 + \xi_{k,j} \left( \mathcal{X^{CH}},\mathcal{X}^{DSu},1 \right) } \right) } \notag \\
&+ \sum\limits_{j \in \mathcal{X}^{CS} \cap \mathcal{D}} w_j \log \Big(1 + \min \big( \xi_{k,j} \left( \mathcal{X^{CH}},\mathcal{X}^{DSu},1 \right), \notag  \\
& \xi_{d,j} \left( \mathcal{X^{CH}},\mathcal{X}^{DSd},1 \right) \big) \Big)  \notag 
\end{align}
\begin{align}
&+ \sum\limits_{j \in \mathcal{X}^{DSu}} {w_j \log \left( {1 + \xi_{k,j} \left( \mathcal{X^{CH}},\mathcal{X}^{DSu},0 \right) } \right) } \notag \\
&+ \sum\limits_{j \in \mathcal{X}^{DSd}} {w_j \log \left( {1 + \xi_{d,j} \left( \mathcal{X^{CH}},\mathcal{X}^{DSd},0 \right) } \right) }. \label{U}
\end{align}
Constraints (\ref{rec}c-f) guarantee the minimum SINR requirements. Constraint (\ref{rec}g) ensures that cellular links and cellular mode D2D links cannot share the same channel. Constraints (\ref{rec}h) and (\ref{rec}i) guarantee the priority of cellular links, and constraint (\ref{rec}j) implies that downlink cellular links cannot access the uplink channel. Then, the optimal channel assignment at the $k$-th stage, $1 \leqslant k \leqslant M_u$, associated with state $\left\{\mathcal{J} \subseteq \mathcal{S}, \mathcal{Z} \subseteq \mathcal{CH}_d \right\}$ can be obtained by
\begin{align}
&\left(\rho_{k,\mathcal{J},\mathcal{Z}} \right)_{i,j}=\\
&\begin{cases}
\left(\rho_{k-1,\mathcal{J}-\mathcal{X}^{All\star}_k,\mathcal{Z}-\{d_k^{\star}\} } \right)_{i,j} &  i < k,\\
                1& i = k,j \in \mathcal{X}^{CS\star}_k \cup \mathcal{X}^{DSu\star}_k, \\
                1& i = d_k^{\star} \ne 0,j \in \mathcal{X}^{CS\star}_k \cup \mathcal{X}^{DSd\star}_k, \\
                0&  i = k, j \notin \mathcal{X}^{CS\star}_k \cup \mathcal{X}^{DSu\star}_k, \\
                0&  i = d_k^{\star} \ne 0, j \notin \mathcal{X}^{CS\star}_k \cup \mathcal{X}^{DSd\star}_k,
\end{cases} \label{rho_cdp}
\end{align}
where $i \in \mathcal{CH}$, $j \in \mathcal{S}$, $\mathcal{X}^\star_k=\{\mathcal{X}_k^{CS\star}, \mathcal{X}_k^{DSu\star}, d_k^{\star}, \mathcal{X}_k^{DSd\star}\} = \mathop {\arg \max } \limits_{ \mathcal{X} } \big[ U \left( k,\mathcal{X} \right)
+ OPT_{k-1, \mathcal{J} - \mathcal{X}^{All}, \mathcal{Z}-\mathcal{X}^{CH}} \big],$
and $\mathcal{X}^{All\star}_k=\mathcal{X}_k^{CS\star} \cup \mathcal{X}_k^{DSu\star} \cup \mathcal{X}_k^{DSd\star}$.
Similarly, the optimal mode selection at the $k$-th stage, $1 \leqslant k \leqslant M_u$, associated with state $\left\{\mathcal{J} \subseteq \mathcal{S}, \mathcal{Z} \subseteq \mathcal{CH}_d \right\}$ can be given by
\begin{align}
&\left(x_{k,\mathcal{J},\mathcal{Z}} \right)_{j}= \notag \\
&\begin{cases}
\left(x_{k-1,\mathcal{J}-\mathcal{X}^{S\star}_k,\mathcal{Z}-\{d_k^{\star}\} } \right)_{j} &  j \in \mathcal{J}-\mathcal{X}^{S\star}_k,\\
                1& j \in \mathcal{X}_k^{CS\star}, \\
                0& j \in \mathcal{X}_k^{DSu\star} \cup \mathcal{X}_k^{DSd\star}.
\end{cases} \label{x_cdp}
\end{align}
 To summarize, the optimum value and optimal channel assignment at the $k$-th stage associated with state $\{\mathcal{J},\mathcal{Z}\} $ can be obtained by the recursive algorithm listed in Algorithm \ref{alg:dp}. The optimal solution at the $M_u$-th stage associated with state $\{\mathcal{S},\mathcal{CH}_d\}$ is the optimal channel assignment for the problem formulated in (\ref{problem}).
\begin{algorithm}[!t]
\caption{Optimal channel assignment and mode selection at the $k$-th stage associated with state $\{\mathcal{J}, \mathcal{Z} \}$}
\label{alg:dp}
\begin{algorithmic}
\IF{$k=0$}
\STATE {Run the DP algorithm in \cite{conf}.}
\ELSE
\STATE {Set $OPT_{k,\mathcal{J},\mathcal{Z}}=-\infty$.}
\FOR {all possible selections of $\mathcal{X}$}
\IF{the $(k-1)$-th stage associated with state $\{\mathcal{J}-\mathcal{X}^{All},\mathcal{Z}-\{d\}\}$ has not been visited}
\STATE Find $OPT_{k-1, \mathcal{J} - \mathcal{X}^{All},\mathcal{Z}-\{d\}} $ and the corresponding channel assignment and mode selection.
\ENDIF
\IF{$OPT_{k,\mathcal{J},\mathcal{Z}} <  U \left( k,\mathcal{X} \right) + OPT_{k-1, \mathcal{J} -\mathcal{X}^{All},\mathcal{Z}-\{d\}}$}
\STATE Set $\mathcal{X}^\star_k = \mathcal{X}$ and $OPT_{k,\mathcal{J},\mathcal{Z}} = U \left( k,\mathcal{X} \right) + OPT_{k-1, \mathcal{J} -\mathcal{X}^{All},\mathcal{Z}-\{d\}}$.
\ENDIF
\ENDFOR
\STATE Update $\left(\rho_{k,\mathcal{J},\mathcal{Z}} \right)_{i,j}$ and $\left(x_{k,\mathcal{J},\mathcal{Z}} \right)_{j}$ as (\ref{rho_cdp}) and (\ref{x_cdp}).
\ENDIF
\end{algorithmic}
\end{algorithm}

\subsection{Complexity Analysis}
The upper bound for time complexity of our proposed DP algorithm can be derived as
\begin{equation} \label{tc_dpall}
\begin{aligned}
&\mathcal{T}_{optimal} \leqslant \\
&
\begin{cases}
\mathcal{O} \Big( MN_d 2^{N_d}\Big) & \text{if } M \leqslant 1,\\
\mathcal{O} \Big( M_{d} N_{dc} N_d  2^{N_{dc}} 3^{N_d}  \Big) & \text{if } M_u=0, M_d >1,\\
\mathcal{O} \Big( M_{u} N_{uc}  N_d 2^{N_{uc}} 3^{N_d}  \Big) & \text{if } M_d=0, M_u >1,\\
\mathcal{O} \Big( M^2_{d} N_c N_{d} 2^{N_{dc}} 3^{N_d}  \Big) & \text{if } M_u=1, M_d\geqslant 1,\\
\begin{array}{l}
\mathcal{O} \Big(  M_u M_d N_d^2 N_{uc} 2^{N_{uc}} 2^{M_d} 4^{N_d}\\
+ N_{dc} N_d 2^{N_{dc}} 2^{M_d} 3^{N_d} \Big)
\end{array} & \text{otherwise},
\end{cases}
\end{aligned}
\end{equation}
where the time complexity of a constant number of products is treated as $\mathcal{O}(1)$ and the constant coefficients and lower order terms are ignored. Considering the space limitation, we omit the proof of (\ref{tc_dpall}). Compared to the time complexity of the exhaustive search approach, which is
$\mathcal{T}_{search} = \mathcal{O} \Big( \sum_{x=0}^{\min(M_u-N_{uc},M_d-N_{dc})} \binom{N_d}{x} \frac{M_{u}!}{\left( M_u-N_{uc}-x \right)!} \cdot \frac{M_{d}!}{\left( M_d-N_{dc}-x \right)!}$  $\times {N_d} (M+1)^{N_d-x}\Big)$,
we can find that our proposed optimal algorithm performs similarly with exhaustive search when the number of channels $M \leqslant 3$ and performs much more efficiently when $M>3$. The proposed DP algorithm can serve as the performance benchmark for other algorithms, even though it has an exponential complexity. As will be shown in the Section V, it can support up to 8 D2D links with 6 channels, which cannot be handled by exhaustive search.
There is one limitation in the DP algorithm, which is that it requires a large memory space given by $\mathcal{O} \Big( M_u 2^{M_d} 2^{\max \left( N_{uc}, N_{dc}\right)} 2^{N_d} \left(N_c+N_d\right)\Big)$.

\section{sub-optimal joint mode selection and channel assignment}
In this section, we propose a practical sub-optimal algorithm. The algorithm jointly optimizes the mode selection and channel assignment via a greedy approach.
\subsection{Bipartite Graph Based Greedy Algorithm}

Since the objective is the weighted sum-rate, the link with a higher weighted rate has a higher priority to be served. Algorithm \ref{alg:cluster} shows the greedy algorithm, where $\mathcal{U}=\mathcal{S} - \bigcup \limits_{i \in \mathcal{CH}} \mathcal{L}^C_i - \bigcup \limits_{i \in \mathcal{CH}} \mathcal{L}^D_i$ denotes the set of links that have not been assigned any channel. To guarantee the priority of cellular links, we need to first find the channel assignment for the cellular links and then find the mode selection and channel assignment for the D2D links. In the following, we will explain how the greedy algorithm works.

To achieve the channel assignment for the cellular links, constraints (\ref{problem}a), (\ref{problem}c), (\ref{problem}d) and (\ref{problem}g) should be guaranteed. Thus, the cellular links should meet their SNR requirements, and each channel can be assigned to at most one cellular link. In addition, uplink cellular links cannot access the downlink channels and downlink cellular links cannot access the uplink channels. Thus, the weighted rate gain of cellular link $j \in \mathcal{C}$ accessing channel $i$ can be given as
\begin{equation}
T^c_{i,j}=
\begin{cases}
-\infty & \text{ if } j \in \mathcal{C}_u \text{ and } i \in \mathcal{M}_d,\\
-\infty & \text{ if } j \in \mathcal{C}_d \text{ and } i \in \mathcal{M}_u,\\
-\infty & \text{ if } {\frac{p_j h_{i,j} }{\sigma^2 }} < \xi^{\text{min}}_{j},\\
w_j \log \left( {1 + \frac{ p_j h_{i,j} }{\sigma^2 }  } \right)& \text{ otherwise}.
\end{cases}
\end{equation}
By regarding the $M$ channels and $N_c$ cellular links as two groups of vertexes, and $T^c_{i,j}$ as the weight of edge between channel $i$ and cellular link $j$, a bipartite graph can be built. The channel assignment for cellular links can be found by solving the maximum weighted bipartite matching problem given by
\begin{align}
\mathop {\max }\limits_{\beta^c_{i,j} } &{ {\sum\limits_{i \in \mathcal{CH},j \in \mathcal{C}} {\beta^c_{i,j}T^c_{i,j}  } } }, \label{problem_c}\\
\text{s.t. } &\sum\limits_{j \in \mathcal{C}} {\beta^c _{i,j} } \leqslant 1, \beta^c_{i,j} \in \{0,1\}, \forall i \in \mathcal{CH},\tag{\ref{problem_c}a}\\
&\sum\limits_{i \in \mathcal{CH}} {\beta^c _{i,j} } \leqslant 1, \beta^c_{i,j} \in \{0,1\}, \forall j \in \mathcal{C},\tag{\ref{problem_c}b}
\end{align}
where $\beta^c _{i,j}=1$ denotes that cellular link $j$ accesses channel $i$. The maximum weight bipartite matching problem can be solved efficiently by the Kuhn-Munkres (KM) algorithm \cite{KM}.

After finding the channel assignment for the cellular links, we need to find the mode selection and channel assignment for the D2D links. In each iteration, an optimal group $\left(i^\star_u,i^\star_d,j^\star,t^\star\right)$ is selected according to the priority value, denoted as $y_{i_u,i_d,j,t}$. The following definitions can help to define the priority value $y_{i_u,i_d,j,t}$.

\begin{defn}
The \emph{channel value} of a channel $i$, denoted as $v^c_{i}(\mathcal{L}_i^C,\mathcal{L}_i^D)$, is the weighted sum-rate of links using channel $i$, which can be expressed as
\begin{equation}
v^c_{i}(\mathcal{L}_i^C,\mathcal{L}_i^D) = \sum\limits_{j \in \mathcal{C} \cap \mathcal{L}_i^C} w_j R_{i,j} \left(\mathcal{L}_i^C,\mathcal{L}_i^D \right)
\end{equation}
where
\begin{align}
&R_{i,j}(\mathcal{L}_i^C,\mathcal{L}_i^D) = \notag \\
&\left\{
\begin{array}{ll}
\log \left( {1 + \xi_{i,j} \left({\mathcal{L}_i^C,\mathcal{L}_i^D,1}\right) } \right) & j \in \mathcal{C} \cap \mathcal{L}_i^C\\
\log \left( {1 + \xi_{i,j} \left({\mathcal{L}_i^C,\mathcal{L}_i^D,0}\right) } \right) & j \in \mathcal{L}_i^D\\
\log \bigg( \min \Big( \xi_{i,j} \left({\mathcal{L}_i^C,\mathcal{L}_i^D,1} \right), & \multirow{2}{*}{$j \in \mathcal{D} \cap \mathcal{L}_i^C$}\\
\sum \limits_{i' \in \mathcal{CH}, i' \ne i} \rho_{i'j} \xi_{i,j} \left(\mathcal{L}_i^C,\mathcal{L}_i^D,1 \right) \Big) \bigg)
\end{array}
\right. \label{rate}
\end{align}
\end{defn}

\begin{defn}
The \emph{priority value} of a D2D link $j \in \mathcal{U}$ while accessing to uplink channel $i_u \in \mathcal{CH}_u \cup \{0\}$ and downlink channel $i_d \in \mathcal{CH}_d \cup \{0\}$ with mode $t$, denoted as $y_{i_u,i_d,j,t}$, equals the channel value gain if it satisfies the constraints (\ref{problem}a), (\ref{problem}b) and (\ref{problem}c), and is $- \infty$ otherwise. Note that, $i_u=0$ means that the D2D link $j$ does not access any uplink channel, and it is similar for $i_d=0$. Moreover, the D2D link $j$ works in the D2D mode and accesses one channel while $t=0$, and works in the cellular mode and accesses one uplink channel and one downlink channel while $t=1$. Thus, $y_{i_u,i_d,j,t}$ is defined as
\begin{align}
&y_{i_u,i_d,j,t}=\notag \\
&\left\{
\begin{array}{ll}
\multirow{3}{*}{\minitab[c]{$v^c_{i_u}(\mathcal{L}_{i_u}^C,\mathcal{L}_{i_u}^D \cup \{j\})$ \\ $-v^c_{i_u}(\mathcal{L}_{i_u}^C,\mathcal{L}_{i_u}^D)$}} & \text{if } t=0,i_u>0,i_d=0, \\& R_{i_u,z}(\mathcal{L}_{i_u}^C,\mathcal{L}_{i_u}^D \cup \{j\}) \geqslant \xi^{\text{min}}_{z} \\& \forall z \in \mathcal{L}_{i_u}^C \cup \mathcal{L}_{i_u}^D \cup \{j\},
\\
\multirow{3}{*}{\minitab[c]{$v^c_{i_d}(\mathcal{L}_{i_d}^C,\mathcal{L}_{i_d}^D \cup \{j\})$ \\ $-v^c_{i_d}(\mathcal{L}_{i_d}^C,\mathcal{L}_{i_d}^D)$}} & \text{if } t=0,i_u=0,i_d>0, \\& R_{{i_d},z}(\mathcal{L}_{i_d}^C,\mathcal{L}_{i_d}^D \cup \{j\}) \geqslant \xi^{\text{min}}_{z} \\& \forall z \in \mathcal{L}_{i_d}^C \cup \mathcal{L}_{i_d}^D \cup \{j\},
\\
\xi \Big( v^c_{i_u}(\{j\},\mathcal{L}_{i_u}^D) & \text{if } t=1,i_u>0,i_d>0, \\
+v^c_{i_d}(\{j\},\mathcal{L}_{i_d}^D) & \mathcal{L}_{i_u}^C=\mathcal{L}_{i_d}^C=\varnothing, \\
-R_{i_u,j}(\{j\},\mathcal{L}_{i_u}^D) \Big) & R_{i_x,z}(\{j\},\mathcal{L}_{i_x}^D) \geqslant \xi^{\text{min}}_{z}\\
& \forall z \in \mathcal{L}_{i_x}^D \cup \{j\}, x \in \{u,d\},
\\
-\infty & \text{otherwise},
\end{array} \label{operator}
\right.
\end{align}
where, $\xi=\min(1,\frac{|U|}{\min (M_u-N_{uc},M_d-N_{dc})})$ is a parameter to adjust the priority for the cellular mode D2D links. Since each cellular mode D2D link occupies two channels while each D2D mode link only accesses one channel. Thus, with the same weighted rate gain, a D2D link has a higher priority to work in the D2D mode.
\end{defn}

\begin{algorithm}[!t]
\caption{Bipartite Graph Based Greedy algorithm}
\label{alg:cluster}
\begin{algorithmic}
\STATE {Set $\rho_{i,j}=0$, $i \in \mathcal{CH}$ and $j \in \mathcal{S}$.}
\STATE {Set $x_{j}=1$, for $j \in \mathcal{C}$, and $x_{j}=0$, for $j \in \mathcal{D}$.}
\STATE {Find the channel assignment for the cellular links by solving problem (\ref{problem_c}) and update $\rho_{i,j}$, $i \in \mathcal{CH}$ and $j \in \mathcal{C}$.}
\STATE {Update the priority value $y_{i_u,i_d,j,t}$, $i_u \in \mathcal{CH}_u$, $i_d \in \mathcal{CH}_d$, $j \in \mathcal{U}$ and $t \in \{0,1\}$.}
\WHILE{$\mathcal{U} \ne \varnothing$ and $\max \limits_{
\begin{subarray} {c}
{i_u \in \mathcal{CH}_u, i_d \in \mathcal{CH}_d}\\
{j \in \mathcal{U}, t \in \{0,1\}}
\end{subarray}
} y_{i_u,i_d,j,t} >0 $}
\STATE {Select $(i_u^\star,i_d^\star,j^\star,t^\star)=\arg \max \limits_{
\begin{subarray} {c}
{i_u \in \mathcal{CH}_u, i_d \in \mathcal{CH}_d}\\
{j \in \mathcal{U}, t \in \{0,1\}}
\end{subarray}
} y_{i_u,i_d,j,t} $.}
\IF {$t^*=0$}
\STATE {Set $\rho_{ \max \{i^\star_u,i^\star_d\},j^\star}=1$ and $x_{j\star}=0$.}
\STATE {Update the priority value $y_{\max \{i^\star_u,i^\star_d\},j,t}$, $j \in \mathcal{U}$ and $t \in \{0,1\}$.}
\ELSE
\STATE {Set $\rho_{ i^\star_u,j^\star}=1$,$\rho_{ i^\star_d,j^\star}=1$ and $x_{j\star}=1$.}
\STATE {Update the priority value $y_{i^\star_u,j,t}$ and $y_{i^\star_d,j,t}$, $j \in \mathcal{U}$ and $t \in \{0,1\}$.}
\ENDIF
\ENDWHILE
\end{algorithmic}
\end{algorithm}

\subsection{Complexity Analysis}

As shown in Algorithm \ref{alg:cluster}, the time complexity of the worst case is $\mathcal{T}_{suboptimal}=\mathcal{O} \left( M^3+M_uM_dN_d^2 \right)$ where constant coefficients and lower order terms are ignored.

\section{simulation results}

Simulation results will be provided to evaluate the performance of our proposed algorithms in this section. We consider a single cell scenario, where all the cellular users are uniformly distributed in the cell, and D2D links are underlaying the cellular network. Same as \cite{conf}, the cluster distribution model in \cite{D2Dmodel} is applied. In the downlink transmission, we adopt an equal power allocation. Small-scale fading, shadow fading and path loss are considered in the channel gains. For instance, the interference channel gain from D2D mode link $z$ to D2D mode link $j$ using channel $i$ is denoted as $h^D_{i,z,j}=K \beta^D_{i,z,j} \zeta^D_{z,j} \left( d_{z,j} \right) ^{-\alpha}$,
where $\beta^D_{i,z,j}$ represents the small scale fading gains, $\zeta^D_{z,j}$ implies the shadow fading gains, $K$ denotes the path loss constant, $d_{z,j}$ is the distance between the receiver of link $j$ and the transmitter of link $z$, and  $\alpha$ denotes the path loss exponent. The small scale fading gains and shadow fading gains are assumed to be independent among all the links and all the channels. The default simulation parameters are the same as \cite{conf}.

\begin{figure}[!t]
  \centering
  \includegraphics[width=2.9in]{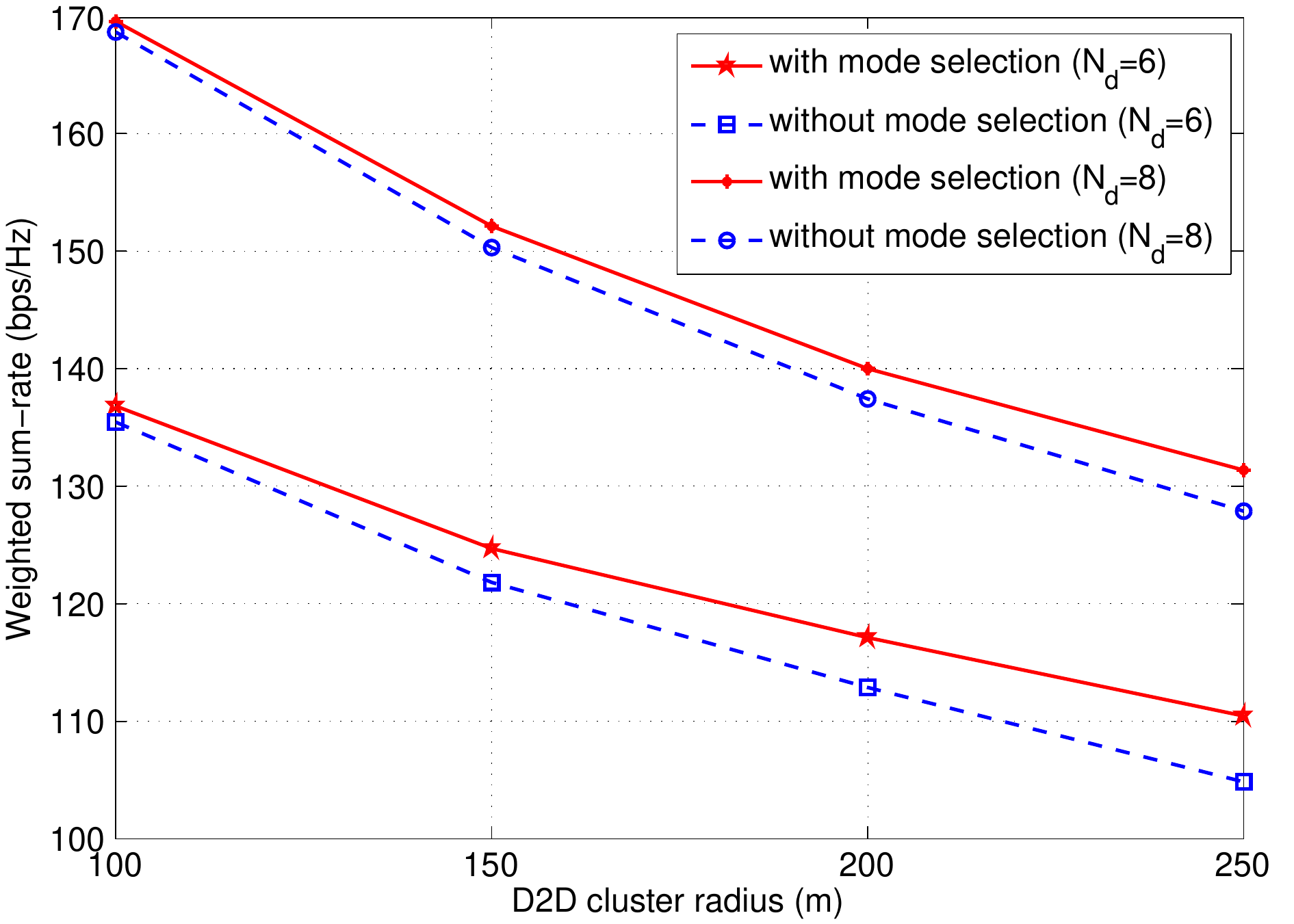}
  \caption{Comparison of the weighted sum-rate for different algorithms with $M_u=M_d=3$ and $N_{uc}=N_{dc}=1$.}
  \label{fig_scheme}
\end{figure}

\begin{figure}[!t]
  \centering
  \includegraphics[width=2.9in]{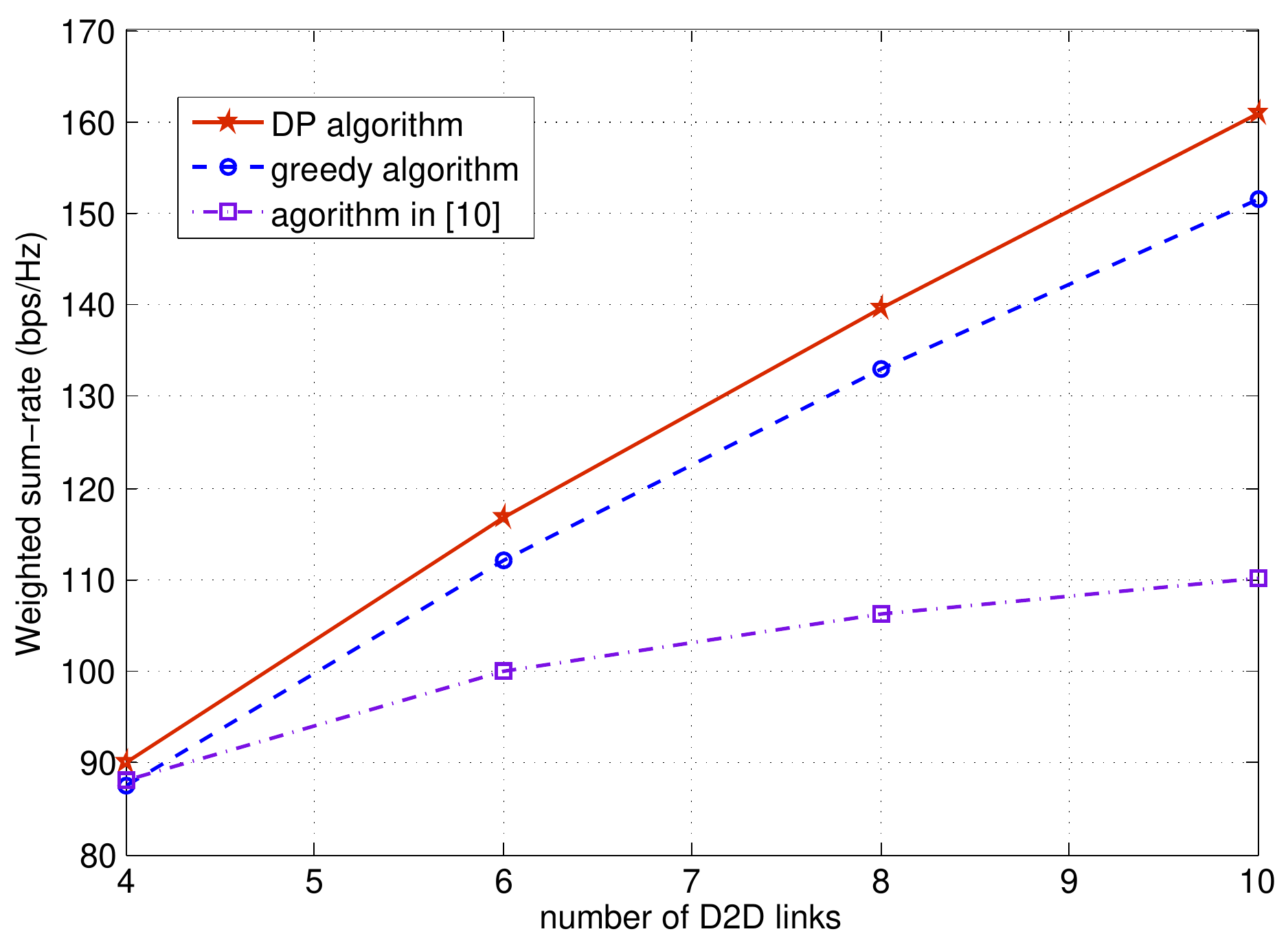}
  \caption{Comparison of sum-rate for different algorithms with $M_u=M_d=2$, $N_{uc}=N_{dc}=1$ and D2D cluster radius equals 150m.}
  \label{fig_scheme2}
\end{figure}

Fig. \ref{fig_scheme} compares the joint mode selection and channel assignment scheme with the one with channel assignment only as considered in \cite{conf}. The optimal DP algorithm is applied in the two scenarios. It is shown that, jointly optimizing the mode selection and channel assignment can improve the weighted sum-rate. Moreover, the gap between the two schemes becomes larger when the distance between D2D users becomes larger, which implies that D2D links are more likely to operate in the cellular mode when the distance between D2D users are relatively large. The performance of our proposed sub-optimal algorithm is validated in Fig. \ref{fig_scheme2} by comparing with the optimal DP algorithm and the previous study \cite{one2onemode}, where at most one cellular link and one D2D link can share the same channel. As shown in Fig. \ref{fig_scheme2}, our proposed bipartite graph based greedy algorithm can provide near optimal performance, and it significantly outperforms the algorithm in \cite{one2onemode}. Moreover, the gap between our proposed algorithm and the algorithm in \cite{one2onemode} becomes larger when the D2D network becomes denser, which demonstrates the superiority of allowing multiple D2D links to access the same channel.


\section{conclusions}
In this paper, we investigated the joint mode selection and channel assignment problem in a cellular network with underlaying D2D communications, where more than one D2D
links may access the same channel. Each link is assumed to have a minimum QoS requirement. An optimal DP algorithm and a bipartite graph based greedy algorithm were proposed. Simulation results validated that our proposed sub-optimal algorithm provides comparable performance to the optimal algorithm. Furthermore, the advantage of jointly optimizing mode selection and channel assignment, as well as the advantage of allowing multiple D2D links to share the same channel, was also demonstrated.

\bibliographystyle{IEEEtran}
\bibliography{IEEEabrv,report}
\end{document}